\documentclass[12pt]{article}
\usepackage{epsfig}

\newcommand{\dslash}{\partial \hskip -0.6em /} 
\newcommand{\fract}[2]{{\textstyle\frac{#1}{#2}}} 

\begin{document}

\baselineskip24pt
\begin{center}
{\Large\bf Casimir Energies in the Light of Renormalizable Quantum 
Field Theories\footnote{Talk presented at the $2^{\rm nd}$ 
intl. workshop {\it Effective Theories of Low Energy QCD}.}}

\vskip1cm

H. Weigel\footnote{Heisenberg--Fellow}\\

\bigskip
Institute for Theoretical Physics, 
T\"ubingen University\\ 
Auf der Morgenstelle 14, 
D--72076 T\"ubingen, Germany

\vskip1cm

{\bf Abstract}
\bigskip

\parbox[t]{15cm}{\small
Effective hadron models commonly require the computation of 
functional determinants. In the static case these are one--loop 
vacuum polarization energies, known as Casimir energies. In this
talk I will present general methods to efficiently compute renormalized 
one--loop vacuum polarization energies and energy densities and apply 
these methods to construct soliton solutions within a variational 
approach. This calculational method is particularly useful to study 
singular limits that emerge in the discussion of the {\it classical} 
Casimir problem which is  usually posed as the response of a fluctuating
quantum field to externally imposed boundary conditions.}

\end{center}

\vskip1cm
\baselineskip14pt

\section*{Introduction}

Many models in hadron physics originate from integrating out 
the more fundamental degrees of freedom like, for example, quarks. 
This change of field variables requires efficient tools to compute 
functional determinants like 
\begin{equation}
{\rm Det}\,\left(\, i\dslash +\Gamma_i\Phi_i -m\,\right)\,,
\label{funcdet}
\end{equation}
in the
Nambu--Jona--Lasinio model~\cite{Na61}
for quark flavor dynamics~\cite{Eb86}. Here $\Phi_i$ 
denote background fields that couple to the internal 
symmetries via the generators $\Gamma_i$. 
Functional determinants are highly ultraviolet divergent
and thus need to be regularized, and eventually 
renormalized by perturbative counterterms.
This is particularly elaborate when the perturbative
expansions become invalid, as for example for soliton
configurations~\cite{Al96}. These configurations are
usually static, $\Phi_i(x)=\Phi_i(\vec{x\,})$, such that
the determinant becomes proportional to the vacuum 
polarization energy
\begin{equation}
\fract{1}{2}\sum_n \big(\omega_n - \omega_n^{(0)}\big)\,.
\label{evac1}
\end{equation} 
Here $\omega_n$ are the eigenvalues of the single particle
Hamiltonian in the presence of $\Phi_i(\vec{x\,})$ while 
$\omega_n^{(0)}$ denote the eigenvalues in the case that
the $\Phi_i$ assume their vacuum values. 

In this talk I will describe
tools to unambiguously regularize and renormalize
such vacuum polarization energies and energy densities starting from the
energy density operator in quantum
field theory. 

\nopagebreak
This presentation is based on work
with E. Farhi, N. Graham, R. L. Jaffe, V. Khemani, M.~Quandt, 
and M. Scandurra. The publications~\cite{GJW,Gr02a,FGJW,FGJWb,Gr02b}
of these collaborations should be consulted for further details.

\section*{Method}

In quantum field theories energies and energy densities are computed as 
renormalized matrix elements of the energy density operator $\hat{T}_{00}$. 
Here I will present the method to unambiguously compute such a matrix element 
when the fluctuating quantum field is coupled to a classical background.
The method is based on expressing this matrix element in terms of 
a Green's function with appropriate boundary conditions. Then the 
energy density is given by a sum over bound states plus an integral 
over the continuum scattering states. Ample use will be made 
of analytic properties of scattering data especially 
to deform momentum integrals along a cut on the positive imaginary
axis. To regulate the ultraviolet divergences of the theory, which
corresponds to eliminating the contribution associated with the
semi--circle at infinite complex momenta, the leading Born approximations 
to the Green's function are subtracted and later exactly added back 
in as Feynman diagrams. These diagrams are then regularized and
renormalized in ordinary Feynman perturbation theory. 

\subsection*{Formalism}

Consider a static, spherically symmetric background potential
$\sigma=\sigma(r)$ with $r=|\vec{x}|$ in $n$ spatial dimensions.
The symmetric energy density operator for a real scalar field coupled
to $\sigma$ is 
\begin{equation}
\hat{T}_{00}(x)=\fract{1}{2}\left[
\dot\phi^2+\phi\left(-\vec{\nabla}^2+m^2+\sigma(r)\right)\phi\right]
+\fract{1}{4}\vec{\nabla}^2\left(\phi^{2}\right)
\label{H_dense}
\end{equation}
with the spatial derivative term rearranged for later use of the 
Schr\"o\-din\-ger equation to evaluate the expression in brackets. 
The ``vacuum'' is the state $|\Omega\rangle$ of lowest energy in the 
background $\sigma$. The ``trivial vacuum'' is the state $|0\rangle$ 
of lowest energy when $\sigma\equiv0$. The vacuum energy density is 
the renormalized expectation value of $\hat{T}_{00}$ with respect to 
the vacuum $|\Omega\rangle$, 
$\langle\Omega|\hat{T}_{00}(x)|\Omega\rangle_{\rm ren}$, 
which includes the matrix elements of the counterterms.
The energy density only depends on the radial coordinate~$r$
since $\sigma$ spherically symmetric,
\begin{equation}
\epsilon(r) = \frac{2\pi^{n/2}}{\Gamma(\frac{n}{2})} r^{n-1}
\langle\Omega|\hat{T}_{00}(x)|\Omega\rangle_{\rm ren} \,.
\label{erad}
\end{equation}
The wavefunctions factorize in radial functions $\phi_\ell(t,r)$
and angle dependent spherical harmonics. The Fock decomposition for the
radial functions reads 
\begin{eqnarray}
\phi_\ell(t,r)&=& \frac{1}{r^{\frac{n-1}{2}}}
\int_0^\infty \frac{dk}{\sqrt{\pi\omega}}
\left[\psi_\ell(k,r)\,e^{-i\omega t}a_{\ell}(k) +
\psi_\ell^*(k,r)\,e^{i\omega t}a^\dagger_{\ell}(k)\right] 
\nonumber \\*
&&+\frac{1}{r^{\frac{n-1}{2}}}\sum_j
\frac{1}{\scriptstyle{\sqrt{\textstyle{2\omega_{\ell j}}}}}
\left[\psi_{\ell j}(r)\,e^{-i\omega_{\ell j} t}a_{\ell j}
+\psi_{\ell j}(r)\,e^{i\omega_{\ell j} t}a^\dagger_{\ell j}\right]\, .
\label{psi_decomp}
\end{eqnarray}
This decomposition contains scattering states with
$\omega=\scriptscriptstyle{\sqrt{\textstyle{k^2+m^2}}}$ and bound states with
$\omega_{\ell j}=\scriptscriptstyle{\sqrt{\textstyle{m^2-\kappa_{\ell j}^2}}}$.
The total angular momentum assumes integer values $\ell = 0,1,2,\ldots$ 
in all dimensions except for $n=1$, where $\ell=0$ and $1$ only, 
corresponding to the symmetric and antisymmetric channels respectively.
The radial wavefunctions $\psi$ are solutions to the 
Schr\"odinger--like equation
\begin{equation} -\psi'' + \fract{1}{r^2}\left(\nu-\fract{1}{2}\right)
\left(\nu+\fract{1}{2}\right)\,\psi +{{\sigma}}(r)\psi - k^{2}\psi =0
\label{Schroedinger}
\end{equation}
where $\nu=\ell-1+\fract{n}{2}$. In each angular momentum channel, 
the wavefunctions are normalized to satisfy the completeness relation
\begin{equation}
\fract{2}{\pi}\int_0^\infty dk\,
\psi_\ell^*(k,r)\psi_\ell(k,r^\prime)+\sum_j
\psi_{\ell j}(r)\psi_{\ell j}(r^\prime) =\delta(r-r^\prime)\,.
\label{complete}
\end{equation}
Then the standard equal time commutation relations for the quantum 
field $\phi$ yield canonical
commutation relations for the creation and annihilation operators,
$[a_\ell(k),a^\dagger_{\ell^\prime}(k^\prime)] =
\delta(k-k^\prime)\delta_{\ell\ell^\prime}$ and $[a_{\ell j},
a^{\dagger}_{\ell'j'}]=\delta_{jj'}\delta_{\ell\ell'}$. All
other commutators vanish. The vacuum $|\Omega\rangle$ is annihilated by 
all of the $a_{\ell}(k)$ and $a_{\ell j}$.
The matrix element~(\ref{erad}) can now be computed by inserting
eq.~(\ref{psi_decomp}) into eq.~(\ref{H_dense}):
\begin{eqnarray}
\epsilon(r)&=&
\sum_\ell N_\ell\Big[\int_0^\infty \frac{dk}{\pi}\omega
\psi_\ell^*(k,r)\psi_\ell(k,r)+
\sum_j\frac{\omega_j}{2} \psi_{\ell j}(r)^2\Big]
\label{ebare} \\
&& + \fract{1}{4}D_r \sum_\ell N_\ell
\Big[\int_0^\infty\frac{dk}{\pi\omega}
\psi_\ell^*(k,r)\psi_\ell(k,r)+
\sum_j\frac{1}{2\omega_{\ell j}}\psi_{\ell j}(r)^2 \Big]
-\epsilon^{(0)}(r)+\epsilon_{\rm CT}(r)\,.
\nonumber
\end{eqnarray}
Here $N_\ell$ is the degeneracy factor, $D_r=\frac{\partial}{\partial
r}\left(\frac{\partial}{\partial r} -\frac{n-1}{r}\right)$,
$\epsilon_{\rm CT}(r)$ is the counterterm contribution, and
$\epsilon^{(0)}(r)$ indicates the subtraction of the energy density in 
the trivial vacuum.

The scattering state contribution is identified with the 
Green's function by defining the local spectral density
\begin{equation}
\rho_\ell(k,r) \equiv \frac{k}{i} G_\ell(r,r,k) \, ,
\label{scatdense1}
\end{equation}
where
\begin{equation}
G_\ell(r,r^\prime,k)=-\fract{2}{\pi}\int_0^\infty dq\,
\frac{\psi_\ell^*(q,r)\psi_\ell(q,r^\prime)}
{(k+i\epsilon)^2-q^2}
-\sum_j\frac{\psi_{\ell j}(r)\psi_{\ell j}(r^\prime)}
{k^2+\kappa_{\ell j}^2}\,,
\label{Green1}
\end{equation}
so that for real $k$
\begin{equation}
\psi_\ell^*(k,r)\psi_\ell(k,r) =
\mathsf{Im} \left\{k\,G_\ell(r,r,k)\right\} =
\mathsf{Re} \left\{\rho_\ell(k,r)\right\} \,.
\label{scatdense2}
\end{equation}
The $i\epsilon$ prescription has been chosen so that this Green's function
is meromorphic in the upper half--plane, with simple poles at the
imaginary momenta $k=i\kappa_{\ell j}$ corresponding to bound states.  
For real $k$, the imaginary part of the Green's function at 
$r=r^\prime$ is an odd function of $k$, while the real part is even. 
Hence eq.~(\ref{ebare}) can be expressed as a contour integral in
the upper half--plane.  The contribution from the semi--circular contour 
at large $|k|$ with $\mathsf{Im}(k)\ge0$ must be eliminated.  
Subtracting sufficiently many terms in the Born series from the Green's
function yields a convergent integral. Then I add back exactly what
I subtracted in the form of Feynman diagrams. The integrand~(\ref{ebare}) 
has a branch cut along the imaginary axis, $k\in [im,+i\infty]$, and simple
poles at the bound state momenta $k=i\kappa_j$. The corresponding
residues cancel the explicit bound state contributions. The discontinuity 
along the cut is hence all what is left to be considered
\begin{eqnarray}
\epsilon(r) &=& -\sum_\ell N_\ell \int_m^\infty\frac{dt}{\pi}
\sqrt{t^2-m^2} \left[1-\frac{1}{4(t^2-m^2)}D_r\right]
\left[\rho_\ell(it,r)\right]_{N}
+ \sum_{i=1}^{N} \epsilon_{\rm FD}^{(i)}(r) + \epsilon_{\rm CT}(r) \cr
&\equiv& \bar\epsilon(r) + \sum_{i=1}^{N} \epsilon_{\rm FD}^{(i)}(r) +
\epsilon_{\rm CT}(r) \, ,
\label{fundamental}
\end{eqnarray}
where
\begin{eqnarray}
\left[\rho_\ell(k,r)\right]_{N} &\equiv&
\rho_\ell(k,r)-\rho_\ell^{(0)}(k,r)-\rho_\ell^{(1)}(k,r)
\ldots -\rho_\ell^{(N)}(k,r) \cr
&=& \fract{k}{i} \left[G_\ell(r,r,k)-G^{(0)}_\ell(r,r,k)
- G^{(1)}_\ell(r,r,k)-\ldots-G^{(N)}_\ell(r,r,k) \right]\,.
\label{state_dens}
\end{eqnarray}
The superscript $(j)$ indicates the term of order $j$ in the
Born expansion. Subtraction of the free Green's function
$G^{(0)}_\ell(r,r,k)$ corresponds to subtracting $\epsilon^{(0)}(r)$
above. The potentially divergent pieces are precisely 
identified~\cite{Schw_Baa,FGHJ} with 
Feynman diagrams $\epsilon_{\rm FD}^{(i)}(r)$,
which are regularized and renormalized using standard methods.  
When combined with the contribution from the counterterms 
$\epsilon_{\rm CT}(r)$ they yield finite contributions to the energy 
density (for smooth backgrounds).

\subsection*{The Radial Green's Function}

A variety of solutions to eq.~(\ref{Schroedinger}) is distinguished
by different boundary conditions:  
\begin{center}
\begin{tabular}{lcl}
free Jost solution &~~~$w_\ell(kr)$:~~~&
$(-1)^\nu\sqrt{\frac{\pi}{2}kr}
\left[J_\nu(kr)+iY_\nu(kr)\right] $\\
Jost solution &~~~$f_\ell(k,r)$:~~~&
$\lim_{r\to\infty} \frac{f_\ell(k,r)}{w_\ell(kr)} = 1 $\\
regular solution &~~~ $\phi_\ell(k,r)$:~~~ &
$\lim_{r\to 0} \frac{\Gamma(\nu+1)}{\sqrt{\pi}}
\left(\frac{r}{2}\right)^{-(\nu+\frac{1}{2})} \phi_\ell(k,r) = 1$ \\
physical scattering solution &~~~$\psi_\ell(k,r)$:~~~&
$\psi_\ell(k,r) = \frac{k^{\nu+\frac{1}{2}}}{F_\ell(k)} \phi_\ell(k,r)$\,.
\end{tabular}
\end{center}
The physical scattering solution is normalized with respect to
the Jost function, $F_\ell(k)$ that is obtained as
the ratio of the interacting and free Jost solutions at
$r=0$,
\begin{equation}
F_\ell(k) = \lim_{r\to 0}\frac{f_\ell(k,r)}{w_\ell(kr)}\,.
\label{Jostfunc}
\end{equation}
In particular, two regular solutions emerge:
$\phi_\ell$ has a simple boundary condition at $r=0$, so that
it is analytic in the upper half $k$--plane; 
$\psi_\ell$ has a physical boundary condition at $r\to\infty$,
corresponding to incoming and outgoing spherical waves.

The Green's function has the simple representation
\begin{equation}
G_\ell(r,r',k) = \frac{\phi_\ell(k,r_{<}) f_\ell(k,r_{>})}{F_\ell(k)}
(-k)^{\nu-\frac{1}{2}}\, ,
\label{Green2}
\end{equation}
where $r_>$ ($r_<$) denotes the larger (smaller) of the two arguments
$r$ and $r^\prime$.  The poles of $G_\ell(r,r^\prime,k)$ occur at the 
zeros of the Jost function, which are the imaginary bound state momenta. 
These are the only poles of eq.~(\ref{Green1}) in the upper
half--plane and, since the two functions in eq.~(\ref{Green1})
and eq.~(\ref{Green2}) obey the same inhomogeneous differential
equation, they are indeed identical.

Although $G_{\ell}$ is analytic in the upper half--plane, $f_\ell$ and
$\phi_\ell$ contain pieces that oscillate for real $k$ and
exponentially increase or decrease when $k$ has an
imaginary part. Actually, only the case $r=r^\prime$ is 
interesting, {\it cf.} eq.~(\ref{state_dens}). Then the product 
$f_\ell\phi_\ell$ is well--behaved. 
This motivates to factorize the dangerous exponential 
components\footnote{For $n=1$ and $n=2$, the case of $\ell=0$ is 
somewhat different~\cite{Gr02a}.}, 
\begin{equation}
f_\ell(k,r) \equiv w_\ell(kr)  g_\ell(k,r) 
\quad {\rm and} \quad
\phi_\ell(k,r) \equiv \frac{(-k)^{-\nu+\frac{1}{2}}}{2\nu}
\frac{h_\ell(k,r)}{w_\ell(kr)}\,,
\label{factorw}
\end{equation}
where $w_\ell$ is the free Jost solution introduced above. With
these definitions,
\begin{equation}
G_\ell(r,r,k) = \frac{h_\ell(k,r)
g_\ell(k,r)}{2\nu g_{\ell}(k,0)}\, .
\label{Green3}
\end{equation}
The definition of
$h_\ell$ does \emph{not} just remove the free part. Instead, it 
enforces the cancellation of $w_\ell$ in the Green's function. 
After analytically continuing to $k=it$, $g_{\ell}(it,r)$ obeys
\begin{equation}
g_\ell''(it,r) = 2 t \xi_\ell(t r) g_\ell'(it,r) + \sigma(r) g_\ell(it,r)
\label{ODE1}
\end{equation}
with the boundary conditions
$\lim_{r\to\infty}g_\ell(it,r) = 1$
and $\lim_{r\to\infty}g_\ell'(it,r) = 0.$
A prime indicates a derivative with
respect to the radial coordinate $r$.  Using these boundary
conditions, one integrates the differential equation numerically for
$g_{\ell}(it,r)$, starting at $r=\infty$ and proceeding to
$r=0$. Similarly, $h_{\ell}(it,r)$ obeys
\begin{equation}
h_\ell''(it,r) = - 2 t \xi_\ell(tr) h_\ell'(it,r)+
\left[ \sigma(r) - 2 t^2 \left.\frac{d
\xi_\ell(\tau)}{d\tau}\right|_{\tau = t r}\right] h_\ell(it,r) \,.
\label{ODE2}
\end{equation}
The factors in eq.~(\ref{factorw}) were chosen to yield simple 
boundary conditions: $h_\ell(it,0) = 0$ and $h_\ell'(it,0) = 1$.
The numerical integration for $h$ starts at $r=0$ and 
runs to $r=\infty$. For real $\tau$,
\begin{equation}
\xi_\ell(\tau) \equiv -\frac{d}{d\tau} \ln\left[w_\ell(i\tau)\right]
\label{xi}
\end{equation}
is real with $\lim_{\tau\to\infty}\xi_\ell(\tau)= 1$, so the two
functions $h_\ell(it,r)$ and $g_\ell(it,r)$ are manifestly real. They
are also holomorphic in the upper half $k$--plane and, most
importantly, they are bounded according to
$|g_\ell(k,r)| \le \hbox{const.}$ and
$|h_\ell(k,r)| \le \hbox{const.}\scriptstyle{[\nu r/(1+|k|r)]}$.
Thus the representation of the
partial wave Green's function in terms of $g_{\ell}$ and
$h_{\ell}$ is smooth and numerically tractable on the positive
imaginary axis.  

The computation of the Born series,
eq.~(\ref{state_dens}), is also straightforward in this
formalism.  The solutions to the differential equations
eq.~(\ref{ODE1}) and eq.~(\ref{ODE2}) are expanded about the free 
solutions,
\begin{eqnarray}
g_\ell(it,r)&=&1+g^{(1)}_\ell(it,r)+g^{(2)}_\ell(it,r)+\ldots \cr
h_\ell(it,r)&=&2\nu r I_\nu(tr) K_\nu(tr)+
h^{(1)}_\ell(it,r)+h^{(2)}_\ell(it,r)+\ldots\, ,
\label{expODE}
\end{eqnarray}
where the superscript labels the order of the background potential~${\sigma}$.
The higher order components obey inhomogeneous linear
differential equations such that ${\sigma}$ is the source term for $g^{(1)}$,
${{\sigma}}g^{(1)}$ is the source term for $g^{(2)}$, and so on. 
Substituting these solutions in the expansion of eq.~(\ref{Green3}) 
with respect to the order of the background potential finally yields 
the Born series for the local spectral density 
\begin{equation}
\left[ \rho_{\ell}(it,r)\right]_{N} =
\left[ t\frac{h_\ell(it,r) g_\ell(it,r)}
{2\nu g_{\ell}(it,0)}\right]_{N}\, .
\end{equation}
Thus I have available a computationally robust representation for 
the Born subtracted energy density $\bar{\epsilon}(r)$, 
{\it cf.} eq.~(\ref{fundamental}).

\subsection*{Feynman Diagram Contribution}

To one--loop order, the Feynman diagrams
of interest are generated by expanding
\begin{equation}
\langle0|\hat{T}_{00}(x)|0\rangle \sim\fract{i}{2} {\rm Tr}\,
\left[\hat{T}_x \left(-\partial^2-m^2 -\sigma\right)^{-1}\right]
\label{t00matrix}
\end{equation}
to order $N$ in the background $\sigma$.  Here $\hat T_{x}$ is the
coordinate space operator corresponding to the insertion of the energy
density defined by eq.~(\ref{H_dense}) at the spacetime point $x$, and
the trace includes space--time integration. The Feynman diagrams
are obtained in ordinary perturbation theory, thus the matrix
element in eq.~(\ref{t00matrix}) is evaluated between the trivial 
vacuum state, which is annihilated by the plane wave
annihilation operators.  The energy density operator has pieces of
order $\sigma^0$ and $\sigma^1$: 
$\hat{T}_x=\hat{T}^{(0)}_x+\hat{T}^{(1)}_x$.
The computation of its vacuum matrix element is most conveniently performed 
in momentum space. The relevant matrix elements are
\begin{equation}
\langle k^\prime | \hat{T}^{(0)}_x | k \rangle
=e^{i(k^\prime-k)x}
\left[k^{0\prime} k^{0} +\vec{k}^\prime\cdot\vec{k} +m ^2\right] 
\quad {\rm and} \quad
\langle k^\prime | \hat{T}^{(1)}_x | k \rangle
=\sigma(x)e^{i(k^\prime-k)x}\, .
\label{t00operator}
\end{equation}
Here I will explicitly consider the contributions to 
$\langle0|\hat{T}_{00}(x)|0\rangle$ that are linear in $\sigma$.
The first contribution of this order comes directly from
$\hat{T}_x^{(1)}$ 
\begin{equation}
\fract{i}{2}{\rm Tr}\left(\frac{1}{-\partial^2-m^2} \hat{T}^{(1)}_x \right)=
\fract{i}{2}\sigma(x)\int \frac{d^dk}{(2\pi)^d}\frac{1}{k^2-m^2}\, .
\label{direct}
\end{equation}
This local contribution is ultraviolet divergent for $d=n+1\ge 2$.  
It is canceled identically by the counterterm in the no--tadpole
renormalization scheme.
An additional contribution at order $\sigma$ originates from
$\hat{T}^{(0)}_x$ and the first--order expansion of the propagator,
\begin{equation}
\fract{i}{2}{\rm Tr}\left(\frac{1}{-\partial^2-m^2}
\hat{T}^{(0)}_x \frac{1}{-\partial^2-m^2}\sigma\right)\,.
\label{twopoint1}
\end{equation}
At ${\cal O}(\sigma)$ the renormalized Feynman diagram contribution 
to the energy density becomes
\begin{equation}
\epsilon_{\rm FD}^{(1)}(r) + \epsilon_{\rm CT}(r) = C_d\, r^{n-1}
\int \frac{d^{d-1}q}{(2\pi)^{d-1}} \tilde\sigma(\vec{q})
e^{i\vec{q}\cdot\vec{x}}\int_0^1 d\zeta
\frac{\zeta(1-\zeta){\vec q\,}^2 }
{\left[m^2+\zeta(1-\zeta){\vec q\,}^2\right]^{2-d/2}}\, ,
\label{edensfd}
\end{equation}
with $C_d=2\scriptstyle{\pi^{\frac{d-1}{2}}
\Gamma\left(2-\frac{d}{2}\right)/
[\Gamma(\frac{d-1}{2})(4\pi)^{d/2}]}$
and $\tilde\sigma(q)=2\pi\delta(q^{0})\tilde\sigma(\vec q)$ is the Fourier
transform of the (time independent) background field. This piece is 
finite for $d=n<4$ and does not contribute to the total energy 
because it vanishes when integrated over space. The extension 
to higher order Feynman diagrams is straightforward.  

\subsection*{Total Energy}

The total energy is simply the integrated energy density of
eq.~(\ref{fundamental}),
\begin{equation}
E[\sigma] = \int_0^\infty \epsilon(r) dr \,\,.
\label{denstot}
\end{equation}
Both, the $t$ integral and the sum over channels in 
eq.~(\ref{fundamental}), are absolutely convergent. This is a 
consequence of deforming the momentum integral in the upper--half 
plane\footnote{For real momenta, $k=it$, this amounts to performing 
the momentum integral before the radial integral.}.
Thus the order of integration can be interchanged 
\begin{equation}
E[\sigma] = - \sum\limits_\ell N_\ell \int_m^\infty \frac{dt}{\pi}\,
\sqrt{t^2-m^2}\int_0^\infty dr \,\left[\rho_\ell(it,r)\right]_N + 
\sum_{i=1}^N E_{\rm FD}^{(i)} + E_{\rm CT} \,,
\label{totimag}
\end{equation}
where the total derivative term has integrated to zero.  As already
explained, a sufficient number, $N$, of Born approximations 
to the local spectral density $\rho_\ell(it,r)$ must be subtracted
to render the $t$ integral convergent. These subtractions are then added
back in as the contribution to the total energy from the Feynman diagrams.
Combined with the contribution from the counterterms this gives
a finite result,
\begin{equation}
\sum_{i=1}^N E_{\rm FD}^{(i)}+E_{\rm CT} = \int_0^\infty dr\,
\left[\sum_{i=1}^N \epsilon_{\rm FD}^{(i)}(r) +\epsilon_{\rm CT}(r)\right]\,.
\end{equation}

In practice, $E[\sigma]$ is more
efficiently computed directly from the perturbation series of the
total energy. First, I'd like to recall that
\begin{equation}
2 \int_0^\infty dr\, \left[\rho(it,r)\right]_N = 
\frac{d}{dt}\left[ \ln F_\ell(it)\right]_N 
= \frac{d}{dt}\left[\ln g_\ell(it,0)\right]_N
\label{app2}
\end{equation}
is valid\footnote{In general, the case 
$\mathsf{Re}(t)=\mathsf{Im}(k)=0$ causes 
uncontrollable oscillations at large $r$: for real $k$ 
the integral $\int_0^\infty dr \left[\rho(k,r)\right]_N$ does 
not exist and the integrated local spectral density cannot be
related to the phase shift.} for $\mathsf{Re}(t)>0$, {\it cf.}
Appendix A of Ref.~\cite{Gr02a}. This allows me to write 
\begin{equation}
E[\sigma] =\sum_\ell N_\ell \int_m^\infty \frac{dt}{2\pi}
\frac{t}{\sqrt{t^2-m^2}}
\left[ \beta_\ell(t,0) \right]_{N} + \sum_{i=1}^{N} E_{\rm FD}^{(i)}
+ E_{\rm CT}\,.
\label{evac3}
\end{equation}
The real function $\beta_\ell(t,r) = \ln g_\ell(it,r)$
is determined by the differential equation
\begin{equation}
-\beta_\ell^{\prime\prime}(t,r)-\left[\beta^\prime_\ell(t,r)\right]^2
+2t\xi_\ell(tr)\beta^\prime_\ell(k,r)+\sigma(r) = 0
\label{betadeq}
\end{equation}
with the boundary conditions 
$\lim_{r\to\infty}\beta(t,r)=\lim_{r\to\infty}\beta^\prime(t,r)=0$.
In $\left[ \beta_\ell(t,0) \right]_{N}$, the first
$N$ Born terms must again be subtracted. They are obtained
by iterating the differential
equation (\ref{betadeq}) according to the expansion of $\beta_\ell(t,r)$
in powers of~$\sigma$. 

To make contact with previous work~\cite{GJW,FGJW,FGJWb,FGHJ}, 
\begin{equation}
E[\sigma] = \sum\limits_\ell N_\ell\left[
\int_0^\infty \frac{dk}{2\pi}\sqrt{k^2 + m^2} \frac{d}{dk}
[\delta_\ell(k)]_N + \frac{1}{2}\sum_j \omega_{\ell j}\right]
+ \sum_{i=1}^N E_{\rm FD}^{(i)} +  E_{\rm CT}
\label{previousdelta}
\end{equation}
observe that along the real axis the phase of the Jost function 
is the scattering phase shift,
\begin{equation}
i \ln F_\ell(k) = i\ln |F_\ell(k)| + \delta_\ell(k)\,.
\label{beckenbauer}
\end{equation}
Equations~(\ref{evac3}) and~(\ref{previousdelta}) are proven
identical by first noticing that for real $k$ $|F_\ell(k)|$ and 
$\delta_\ell(k)$ are respectively even and odd functions and 
then computing the momentum integrals along the branch 
cut~ $k\in[im,+i\infty]$.

\section*{Soliton Formation in a D=1+1 Chiral Model} 

The described method is well suited to efficiently and unambiguously 
compute vacuum polarization energies in renormalizable quantum field 
theories. Then the total energy, {\it i.e.} the sum of the classical 
and vacuum polarization energies, is a functional of the background 
field. Varying this background field maps an energy surface. The 
existence of a local minimum on that surface indicates the existence 
of an energetically stable solution to the equation of motion, a 
soliton\footnote{The minimum itself is not necessarily a soliton
because the space of variational parameters is limited and the
exact soliton might have even lower energy.}. Of particular 
interest are models that do not contain soliton solutions at 
the classical level such that solitons get stabilized by quantum
corrections.

Now I would like to consider this idea in the framework of a simple chiral
model in $D=1+1$~\cite{FGJW}. The realistic $D=3+1$ case is more difficult 
and a discussion is presented in Ref.~\cite{FGJWb}.
In this two--dimensional model a two--component boson 
field $\vec{\phi}=(\phi_1,\phi_2)$ couples chirally to a 
fermion $\Psi$ that come in $N_f$ (equivalent) modes:
\begin{eqnarray}
{\cal L}=\frac{1}{2}\, \partial_\mu\vec{\phi}\cdot\partial^\mu\vec{\phi}
+\sum_{n=1}^{N_f}\bar{\Psi}_i\left\{i\partial \hskip -0.5em /
- G\left(\phi_1+i\gamma_5\phi_2\right) \right\}\Psi_i\, .
\label{lagd11}
\end{eqnarray}
where the potential for the boson field
\begin{eqnarray}
V(\vec{\phi})=\frac{\lambda}{8}
\left[\vec{\phi}\cdot\vec{\phi}
-v^2+\frac{2\alpha v^2}{\lambda}\right]^2
-\alpha v^3\left(\phi_1-v\right)+{\rm const.}\,
\label{potential}
\end{eqnarray}
contains a term (proportional to $\alpha$) that breaks the chiral symmetry 
explicitly in order to avoid problems stemming from (unphysical) infra--red 
singularities that occur when the vacuum configuration would be determined 
via the na{\"\i}ve treatment of spontaneous symmetry breaking\cite{Col73}.
In this manner it is guaranteed that the VEV is given by
$\langle \vec{\phi}\rangle=(v,0)$. Here the counterterm Lagrangian is
not presented explicitly. It is determined such that the quantum 
corrections lead to a vanishing tadpole diagram for the boson field.
Considering only the classical contribution 
does \underline{not} support a stable soliton soliton.

In the limit that the number of fermion modes becomes large with 
$v^2/N_f\sim{\cal O}(1)$ only the classical and one fermion loop pieces 
contribute. In the following I will only consider that limit, 
{\it i.e.} $E_{\rm tot}=E_{\rm cl}+E_{\rm F}$. The fermion contribution 
can be split into two pieces $E_{\rm F}=E_{\rm vac}+E_{\rm val}$. 
The valence part $E_{\rm val}$ is given in terms
of the bound state energies such as to saturate the total fermion number
that is fixed to be $N_F$. The vacuum piece is computed according to
the formalism described in the preceding section:
\begin{eqnarray}
E_{\rm vac}[\vec{\phi\,}]=-\frac{1}{2}\sum_i^{\rm b.s.}\left(
\left|\omega_i\right|-Gv\right)
-\int_0^\infty\frac{dk}{2\pi}\, \left(\omega_k-Gv\right)
\frac{d}{dk}\left(\delta_{\rm F}(k)-\delta^{(1)}(k)\right)\, ,
\label{efermion}
\end{eqnarray}
which is obtained from Eq~(\ref{previousdelta}) by employing 
Levinson's theorem. Note the overall ``-'' sign for fermions and
recall that the single particle spectrum is not charge conjugation
invariant. Furthermore,
$\delta_{\rm F}$ denotes the sum of the eigenphase shifts\footnote{The
eigenchannels are labeled by parity and the sign of the single particle 
eigenenergies.}. The subtraction 
\begin{equation}
\delta^{(1)}(k)=\frac{2G^2}{k}\int_0^\infty dx 
\left(v^2-\vec{\phi}\,^2(x)\right)
\label{subtr}
\end{equation}
that renders $E_{\rm vac}$ finite
contains both first and second order Born approximants in the fluctuations
of $\vec{\phi}$ about $\langle \vec{\phi}\rangle$. The first order 
is unambiguously fixed by the no--tadpole renormalization condition and
the second order by the chiral symmetry. 

\begin{figure}[t]
\centerline{
  \epsfig{file=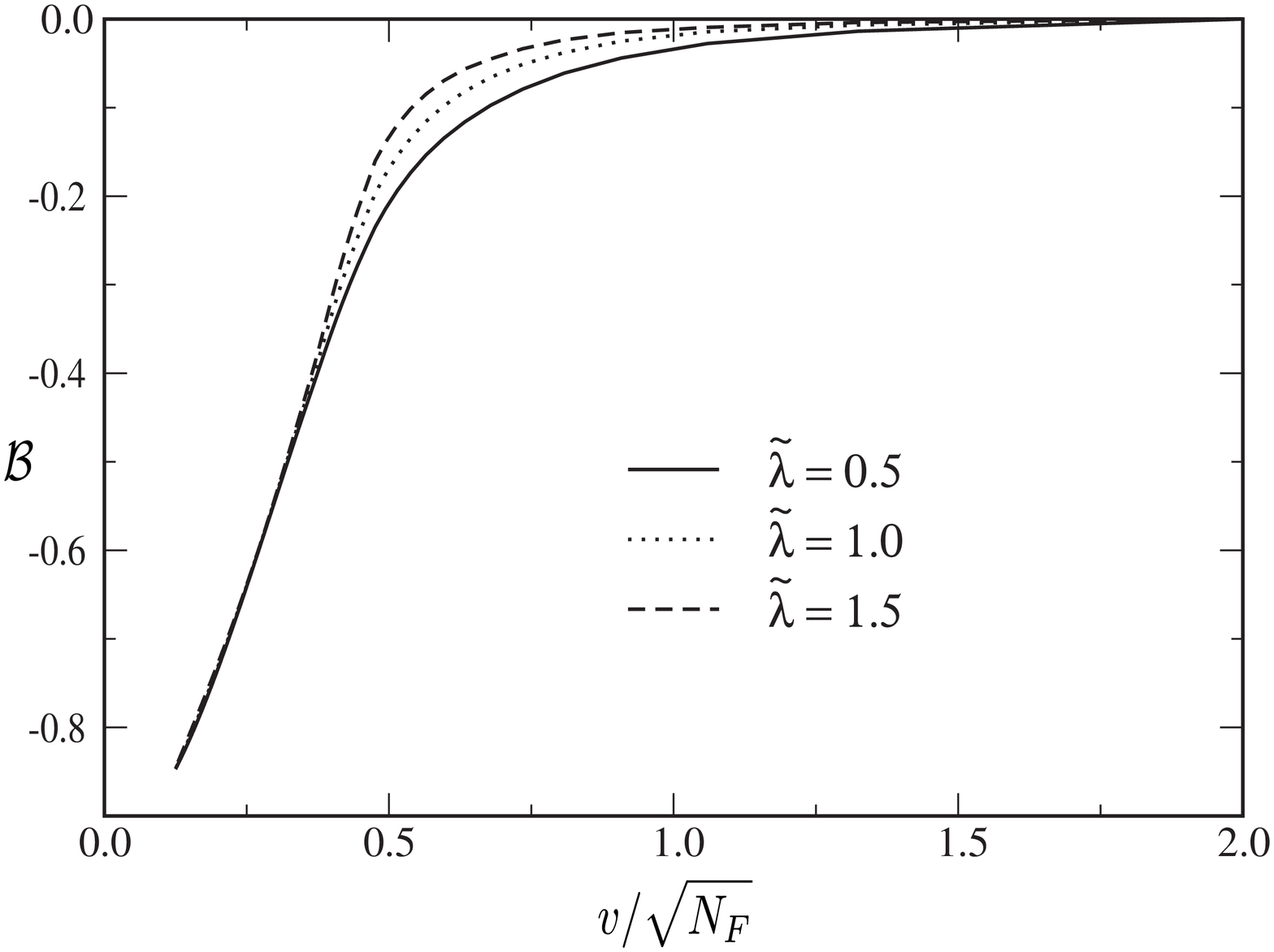,height=4cm,width=7cm}\hskip2cm
  \epsfig{file=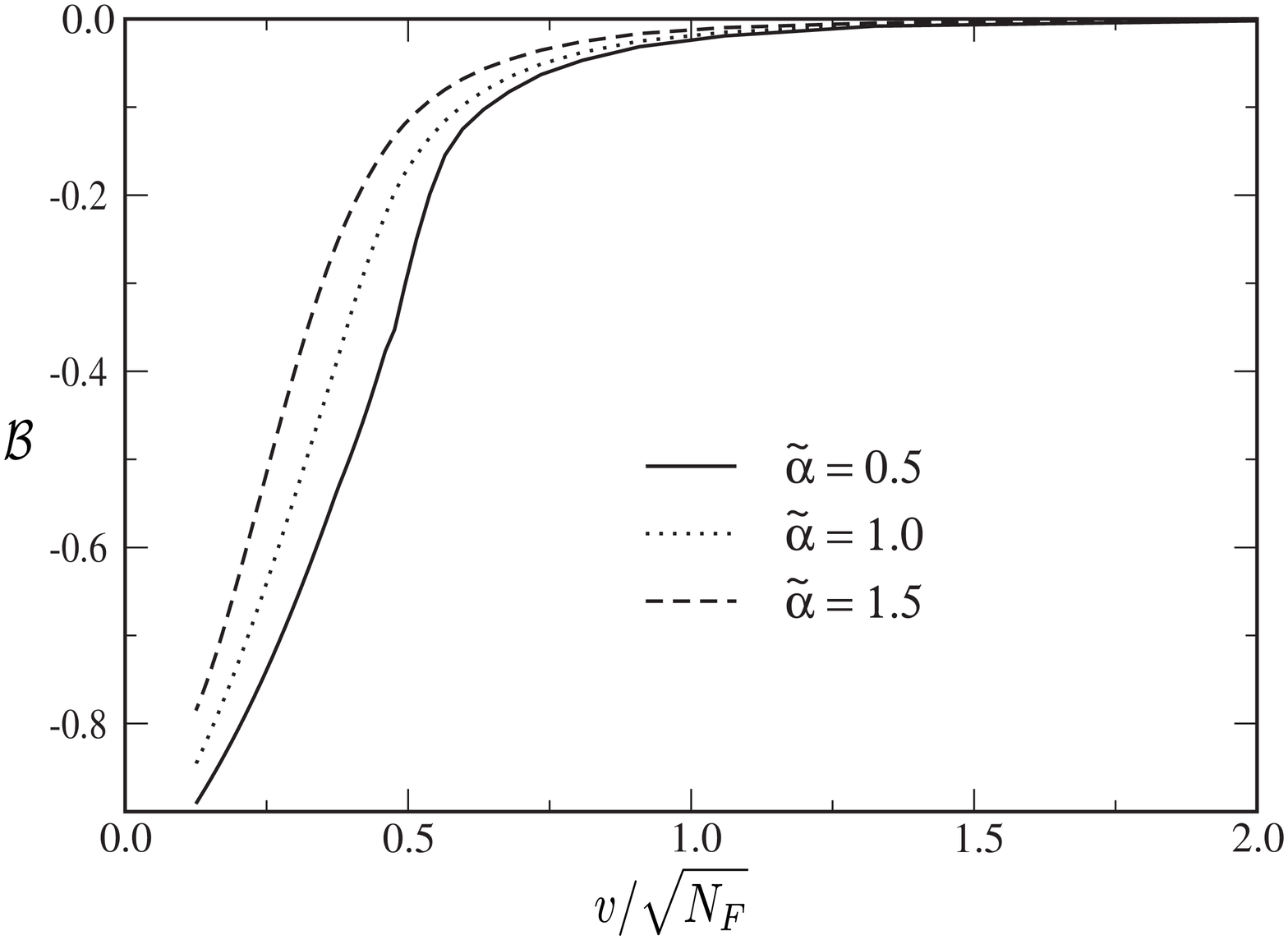,height=4cm,width=7cm}}
\caption{\label{fig_1}\sf\small
The maximal binding energy as a function of the model 
parameters as obtained from the {\it Ansatz}~(\protect\ref{variation}) 
in units of $Gv$; $\tilde{\alpha}={\alpha}/{G^2}$ and
$\tilde{\lambda}={\lambda}/{G^2}$.}
\end{figure}

Having established the energy functional I now consider
variational {\it Ans\"atze} for the background field that turn this
functional in a function of the variational parameters. As an example 
I assume
\begin{equation}
\phi_1+i\phi_2=v\left\{1-R+R\,{\rm exp}\hspace{0.1mm}
\left[i\pi\left(1+{\rm tanh}(Gvx/w)\right)\right]\right\}
\label{variation}
\end{equation}
that introduces width ($w$) and amplitude ($R$) parameters. For
prescribed model parameters ($G$,$v$,etc.) the energy must be
minimized with respect to $w$ and $R$. The resulting binding 
energy ${\cal B}=E_{\rm tot}-Gv$ is shown in Fig.~\ref{fig_1}.
Even though the {\it Ansatz}~(\ref{variation}) may not be the final answer
to the minimalization problem, ${\cal B}$ is definitely negative.
Thus a solitonic configuration is energetically favored showing that 
indeed quantum fluctuations can create a soliton that is not stable at
the classical level.

\section*{Casimir energies}

As noted earlier, the presented method to compute vacuum 
polarization energies is not limited to the case of smooth
background fields. It is particularly interesting to employ 
singular background fields to imitate boundary conditions of the 
fluctuating quantum field. It is known for some time~\cite{Caold} 
that the vacuum polarization energy diverges 
when the fluctuating field is constrained by boundary conditions.
{\it Ad hoc} schemes for their removal have been proposed~\cite{AAA}.
However, the method of renormalization in continuum quantum 
field theory represents the \emph{only} physical way to
treat these divergences.

\subsection*{Dirichlet Points in One Space Dimension}

The simplest example to be considered is that of a massive scalar field
$\phi(t,x)$ in one dimension, constrained to vanish at $x=-a$ and $a$. 
The standard approach, in which the boundary conditions are imposed
{\it a priori\/}, gives an energy \cite{MT}
\begin{equation}
\widetilde{E}_{2}(a) = -\frac{m}{2} -
\frac{2a}{\pi}\int_{m}^{\infty}dt\frac{\sqrt{t^2-m^2}}{e^{4at}-1} \,.
\label{wrongenergy}
\end{equation}
The tilde denotes the imposition of the Dirichlet boundary
condition at the outset.  This expression yields an
attractive force between the two Dirichlet points,
\begin{equation}
\widetilde{F}(a)=-\frac{d\widetilde{E}_{2}}{d(2a)}
=-\int_{m}^{\infty}\frac{dt}{\pi}\frac{t^2}{\sqrt{t^2- m^2}(e^{4at}-1)}\, .
\label{force1}
\end{equation}
In the massless limit it simplifies considerably: 
$\widetilde{E}_{2}(a)=-\pi/48a\,$ and $\,\widetilde{F}(a)=-\pi/96a^2$.
But this result is not internally consistent: as $a\to\infty$,
$\widetilde{E}_{2}(a)\to 0$, indicating that the energy of an isolated
``Dirichlet point'' is zero.  The limit $a\to 0$ also describes 
a single Dirichlet point, but $\widetilde{E}_{2}(a)\to\infty$ as $a\to 0$. 
Also note that $\widetilde{E}_{2}(a)$ is well defined as $m\to 0$, even
though scalar field theories in one space dimension become infrared 
divergent when~$m\to 0$.

The Dirichlet point problem can nicely be studied with the 
presented method. A single delta--function background 
\begin{equation}
\sigma_1=\lambda\delta(x-a)
\label{onedelta}
\end{equation}
has the Green's function
\begin{equation}
G_\lambda(x,y)=G_0(x,y)-\frac{\lambda G_0(x,a)G_0(a,y)}
{1+\lambda G_0(a,a)}
\label{Green1delta}
\end{equation}
where $G_0(x,y)$ is the Green's function in the non--interacting case. 
The momentum argument has been omitted. Obviously the limit 
$\lambda\to\infty$ gives Dirichlet boundary conditions, 
$G_\infty(x,a)=G_\infty(a,y)=0$.
The vacuum polarization energy associated with eq.~(\ref{onedelta})
is~\cite{Gr02a} 
\begin{equation}
    E_{1}(\lambda) =  \int_{m}^{\infty}\frac{dt}{2\pi}\,
    \frac{t\ln\left[1+\frac{\lambda}{2t}\right]
    -\frac{\lambda}{2}}{\sqrt{t^{2}-m^{2}}}
\end{equation}
To study the problem of two Dirichlet points it is 
obvious to consider
\begin{equation}
\sigma_2(x)=\lambda\left[\delta(x+a)+\delta(x-a)\right]\,.
\label{sigmadelta}
\end{equation}
The renormalized Casimir energy for this potential 
has also been computed in Ref.~\cite{Gr02a},
\begin{equation}
E_2(a,\lambda)=\int_m^{\infty}\frac{dt}{2\pi}\frac{1}
{\sqrt{t^2-m^2}}\left\{ t \ln \left[
1+\frac{\lambda}{t}+\frac{\lambda^2} {4t^2}(1-e^{-4at})\right] -
\lambda\right\}
\label{e2}
\end{equation}
For any finite coupling $\lambda$, the inconsistencies noted in
$\widetilde{E}_{2}(a)$ do not afflict $E_{2}(a,\lambda)$: as
$a\to\infty$, $E_{2}(\lambda)\to 2E_{1}(\lambda)$, and as $a\to 0$,
$E_{2}(a,\lambda)\to E_{1}(2\lambda)$.  Also $E_{2}(a,\lambda)$
diverges logarithmically in the limit $m\to 0$ as it should.  The
\emph{force}, obtained by differentiating eq.~(\ref{e2}) with respect to
$2a$, agrees with eq.~(\ref{force1}) in the limit $\lambda \to
\infty$. However $E_{2}(a,\lambda)$ \emph{diverges} like 
$\lambda\log\lambda$ as $\lambda\to\infty$.  Thus the renormalized 
Casimir \emph{energy} diverges as the Dirichlet boundary 
condition is imposed, a physical effect which is missed in the 
pure boundary condition calculation.

\begin{figure}[t]
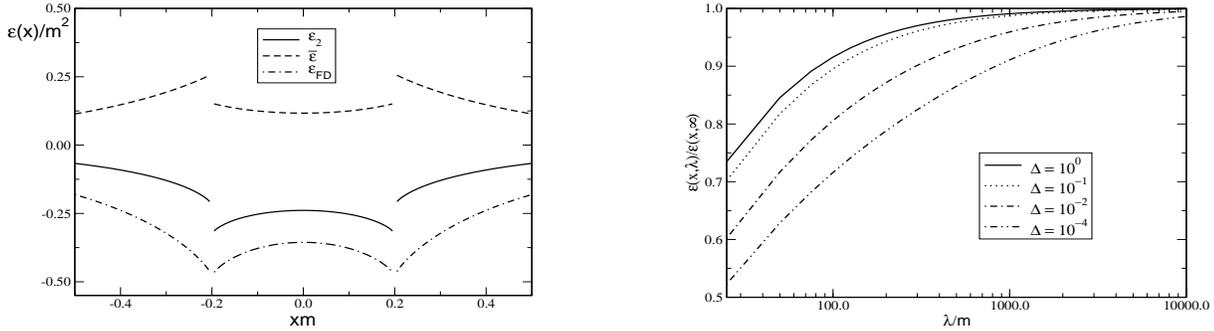

\centerline{
  \epsfig{file=dens1.eps,height=4.3cm,width=7cm}\hskip2cm
  \epsfig{file=nonuniform.eps,height=4.3cm,width=7cm}}
\caption{\label{fig_2}\sf\small The energy density for the two 
delta function background at $a=0.2/m$ computed from 
eqs~(\ref{fundamental}) and~(\ref{edensfd}). Left panel:
$\lambda=3m$. The distinct contributions associated with 
eqns.~(\protect\ref{fundamental}) and~(\protect\ref{edensfd}) are 
disentangled but the singular contributions at $x=\pm a$ are omitted.
Right panel: Dirichlet limit $\lambda\to\infty$ in 
comparison to the the boundary condition 
result~(\protect\ref{edens1}); $\Delta=(a-x)/a$.}
\end{figure}

The counterterms to the energy density vanish away from $x=\pm a$ 
because they are local functions of $\sigma(x)$. Therefore
the Casimir \emph{energy density} for $x\ne\pm a$ can be calculated
assuming Dirichlet boundary conditions from the start simply
by subtracting the density in the absence of boundaries without
encountering any further divergences~\cite{MT},
\begin{eqnarray}
\tilde{\epsilon}_{2}(x,a) & = &
-\frac{m}{8a}- \int_{m}^{\infty}\frac{dt}{\pi}
\frac{\sqrt{t^2-m^2}}{e^{4at} -1} 
-\frac{m^2}{4a}\sum_{n=1}^{\infty}\frac{\cos
\left[\frac{n\pi}{a}(x-a)\right]}{\sqrt{(\frac{n\pi}{2a})^2+m^2}}
\quad
\mbox{for~} |x|<a \nonumber \\
\tilde{\epsilon}_{2}(x,a) & = & -\frac{m^2}{2\pi}K_0(2m|x-a|)
\quad\mbox{for~} |x|>a \,.
\label{edens1}
\end{eqnarray}
This result excludes the points $x=\pm a$. For finite $\lambda$ the 
Casimir energy density, $\epsilon_2(x,a,\lambda)$, was also computed 
in Ref.~\cite{Gr02a} and is displayed in Fig.~\ref{fig_2}.  
The energy density between the isolated points is negative
and approaches the boundary condition limit~(\ref{edens1})
in a non--uniform manner.
In the limit $\lambda\to\infty$ it agrees with
eq.~(\ref{edens1}) except at $x=\pm a$ where it contains extra,
singular contributions.  If one integrates eq.~(\ref{edens1}) over all
$x$, ignoring the singularities at $x=\pm a$, one obtains
eq.~(\ref{wrongenergy}). Including the singular contributions at 
$\pm a$ by integrating $\epsilon_2(x,a,\lambda)$ gives eq.~(\ref{e2}).

This simple example illustrates the principal situation: In the
Dirichlet limit the renormalized Casimir energy diverges because the
energy density on the ``surface'', $x =\pm a$ diverges.  However the
Casimir force and the Casimir energy density for all $x\neq\pm a$
remain finite and equal to the results obtained by imposing the
boundary conditions {\it a priori\/}, eqs.~(\ref{force1}) and
(\ref{edens1}).

\subsection*{Two Space Dimensions}

A scalar field in two space dimensions constrained to vanish on a circle 
of radius $a$ presents a more complex problem. For smooth backgrounds
only the local {\it tadpole} diagram diverges and thus the no--tadpole
renormalization condition is still sufficient to render the theory finite.
For $\sigma(\vec{x\,})=\lambda\delta(r-a)$ and $r\ne a$ 
the subtracted local spectral density $[\rho_{\ell}(it,r)]_0$ 
vanishes \emph{exponentially} as the momentum along the 
cut in eq.~(\ref{fundamental}) increases. For finite $\lambda$, 
both the $t$-integral and the $\ell$-sum are uniformly convergent 
so $\lambda\to\infty$ can be taken under the sum and integral. The 
resulting energy density, $\epsilon(r,\lambda)$, agrees with 
$\tilde\epsilon(r)$, obtained when the Dirichlet
boundary condition, $\phi(a)=0$, is assumed from the start.  As in one
dimension, nothing can be said about the total energy because
$\tilde\epsilon(r)$ is not defined at $r=a$, but unlike the
one dimensional case, the integral $\int dr\,\epsilon(r,\lambda)$ 
now diverges even in the sharp limit for finite $\lambda$.

To understand the situation better, let's consider $\sigma(\vec{x\,})$ to 
be a narrow Gau{\ss}ian of width $w$ centered at $r=a$ and explore the
sharp limit where $w\to 0$ and $\sigma(\vec{x\,})\to\lambda\delta(r-a)$.  
For $w\ne 0$, $\sigma$ does not vanish at any value of $r$, so 
$[\rho_{\ell}(it,r)]_0$ no longer falls exponentially at large 
$t$ (and $r\ne a$), and subtraction of the first Born approximation to
$\rho_{\ell}(it,r)$ is necessary, {\it i.e.} $N=1$ in 
eq.~(\ref{fundamental}). As noted above, the
compensating tadpole graph can be canceled against the counterterm,
$c_{1}\lambda\sigma(\vec{x\,})$.  The result is a renormalized Casimir
energy density, $\epsilon(r,w,\lambda)$, and Casimir energy,
$E(w,\lambda)=\int_{0}^{\infty}\,dr\, \epsilon(r,w,\lambda)$, both of
which are finite.  However as $w\to 0$ both $\epsilon(a,w,\lambda)$
and $E(w,\lambda)$ diverge.

The divergence can be traced to the ${\cal O}(\lambda^2)$ Feynman diagram.  
This diagram is separated by subtracting the \emph{second} Born 
approximation to $\rho_{\ell}(it,r)$, {\it i.e.} $N=2$ in 
eq.~(\ref{fundamental}). Then the  $\ell$-sum and $t$-integral no 
longer diverge in the sharp limit, $w\to0$ but the equivalent 
diagram must be added back explicitly. In the limit $w\to 0$ it contributes
\begin{equation}
- \frac{\lambda^2\,a^2}{8}\,
\int_0^M dp\,J_0^2(a p)\,\arctan\frac{p}{2m}  
\label{B6}
\end{equation}
to the total energy. This
diverges logarithmically as $M\to\infty$. The divergence 
originates from the high momentum components in the Fourier transform of
$\sigma(\vec{x\,})=\lambda\delta(r-a)$ rather than the high energy behavior 
of the loop integral. This divergence gives an infinite contribution to 
the stress because it varies with the radius of
the circle. This divergence only gets worse in higher dimensions
(in contrast to the claim of Ref.~\cite{Milton}).  For example, 
for $\sigma(r)=\lambda\delta(r-a)$ in three space dimensions the
\emph{renormalized} two point function is proportional to
$\lambda^{2}a^{4}\int_0^M dp f(p)$ with $f(p)= p^{2} j_{0}^{2}(pa)\ln p$ 
for large $p$. This integral diverges like $M\ln(M)$. 

The imposed upper limit, $M$, in eq.~(\ref{B6}) plays the role 
of a \emph{physical} cutoff 
that regulates divergences localized on the surface. It
is not related to the regulator of the ultraviolet divergences
in loop integrals. Hence divergences like in eq.~(\ref{B6}) are
not renormalized by standard
counterterms whose (divergent) coefficients are independent of the 
considered background because they are fixed by renormalization 
conditions on Green's functions at some prescribed finite external momenta. 
Divergences that emerge as $M\to\infty$ indicate that even the sharp 
limit $w\to0$ does not exist. The strong coupling limit 
$\lambda\to\infty$ makes the divergence even worse.
If the divergent terms depend on the quantity conjugate
to the force (tension) under consideration\footnote{For example,
the distance between plates or the radius of the shell.}
the force (tension) cannot be defined independently of the 
structure of the material. This is the case for the shell 
or the sphere, but not for rigid bodies.

\section*{Conclusion}

In this talk I have presented an efficient method to compute
vacuum polarization energies in renormalizable quantum
field theories for static background fields. Starting point
for this method is the energy density operator. Its matrix 
element in the ground state is expressed in terms of the 
Green's function which subsequently is parameterized by data
from scattering off the background field. To compute the 
momentum integrals as contour integrals in the upper--half
plane, this approach makes ample use of the identity of Feynman 
diagrams and Born approximants to Casimir energies. More 
importantly, this identity allows one to implement standard 
(perturbative) renormalization conditions on the divergent, low
order Green's functions. In this way the removal of the 
ultraviolet loop divergences is independent of the considered
background.

Utilizing a variational approach to the so--obtained total
energy, solitons can be constructed. As an application I
have shown that in a $1+1$ dimensional chiral model quantum
corrections create a soliton that is classically unstable.
In similar $3+1$ dimensional models the situation is more 
complex~\cite{FGJWb}, as issues like Landau poles~\cite{landau} 
and sphaleron barriers~\cite{sphal} complicate matters.

The divergences that arise when a quantum field is forced to vanish 
on a surface can be nicely studied in this approach by
implementing a boundary condition as the limit of a less singular 
background. Energy densities away from the surfaces or quantities 
like the force between rigid bodies, for which the surfaces can be held 
fixed, are finite and independent of the material cutoffs. Observables 
that require deformation or change in area of the surface cannot be 
defined independently of the other material properties.

\section*{Acknowledgments}
I would like to thank the organizers for this 
interesting and worthwhile workshop. Furthermore I appreciate
helpful remarks on the manuscript by M. Quandt
and acknowledge support by the Deutsche Forschungsgemeinschaft 
(DFG) under contract We~1254/3-2.

\end{document}